# OPTICAL TRANSITIONS IN NEW TRENDS ORGANIC MATERIALS

[1]Pérez-Merchancano S.T. [2]Marques G. E. and [1]Bolivar-Marinez L. E.
[1] Departamento de física, Universidad del Cauca calle 5 # 4-70,
Popayán, Cauca Colombia.
[2] Departamento de Física, Universidade Federal de São Carlos,
13565-905, São Carlos, S. P. Brazil

The PTCDA (3,4,9,10-Perylene-tetracarboxylic dianhydride) and the NTCDA (1,4,5,8-Naphtalenetetracarboxylic dianhydride) are aromatic, stable, planar and highly symmetric with unusual electrical properties. The PTCDA is a semiconductor organic crystalline of particular interest due to its excellent properties and electronic potential that are used in optoelectronic devices and the NTCDA it is monoclinic and its space group is similar to that of the PTCDA. Recently, alternate layers of PTCDA and NTCDA were growth forming multiple structures of quantum wells showing a new class of materials with new optic lineal properties. Some have assured that their big utilities would be centered in the construction of diodes and of possible guides of waves. We have carried out calculations semi-empirical of the electronic structures and of optic properties of the PTCDA and of the NTCDA that show us that they are structures highly orderly polymeric, semiconductors in a negative load state (charge state= -2).

**Introduction**

In the last two decades one has come carrying out considerable efforts in the study of all series of compound organic and inorganic that are material appropriate for electronic applications [1,2]. they have been carried out studies of junctures of material type metal-insulating-semiconductor with the purpose of creating devices (diodes) by means of different deposition techniques [3,4]. We have investigated the electronic, geometric and optical properties of organics materials, as the 3,4,9,10-perylenetetracarboxylic dianhydride and 1,4,5,8-naphtalenetetracarboxylic dianhydride (to see figure 1) [1-3]. These materials are used to fabrication of the organic semiconductors diodes and are extremely easy to fabricate and have reproducible electrical characteristic [3].

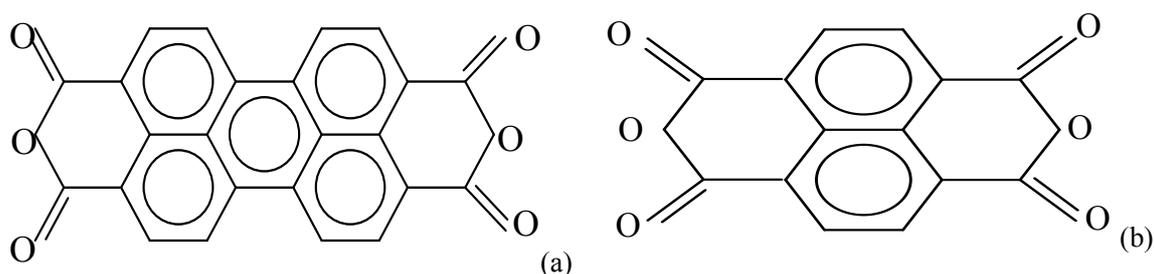

**Figure 1**. Structural configuration of the (a) PTCDA and (b) NTCDA.

Some of these properties, low optic absorption that gets lost in wave longitudes near the infrared, environmental stability and modeling capacity to generate structures of low dimensionality, is those that exactly motivate to investigate its utility. These juncture possesses some important characteristics that they have gotten the attention thoroughly. The used organics materials are an aromatics, stables compounds, easy to manufacture, with characteristic electric reproducible and variety of behaviors, since in its natural state they are highly resistive, but when they are irradiated with faces of great energy they become excellent conductors[3]. Experimentally these materials has been characterized in two ways: first for the width of their forbidden band (gap) that is of the order of 2.2 ± 0.1 eV [4] and it classifies it as a semiconductor, and due to their parallel growth and to that their planar spaces are very near (an extensive overlapping of orbital p) the delocalization of electrons is observed that is in a high intrinsic conductivity (bigger than $10^{-5}$ (Wcm$^{-1}$)[4]. In this document we present a theoretical study of the electronic structure (geometric conformation and some optic properties) of the aromatic compound PTCDA, and NTCDA at the same time we show that their behavior like semiconductor you can fulfill bigger easiness when it is in their ionized state.

**Methodology**

The study and characterization of these polymeric materials one has come developing for two decades. One of the pioneer groups in this study is that of S.R. Forrest [4] who they have determined that the PTCDA compound presents diversity of physical properties with possibility of wide technological application. With the purpose of contributing to understanding and characterization of such compounds we seek in first instance, to carry out the geometric optimization of the PTCDA in their neutral and charge state (±1, ±2), using for it the semi-empiric methods PM3 (Parametric Method 3) [5] of Mopac and Spartan [6], where this last one allows or not to include the molecular structure in a means. This means cannot be a solvent polar (as the water). In the second part of our investigation that is the optic characterization, we will use Zindo-CI (Zerner´s intermediates you Neglect of Differential Overlap) [7]. The whole methodological process before mentioned it has been used with success in the description of organic molecules [5].

**Results and Discussion**

The first table show the results obtained for the head formation with the methods PM3 and SPARTAN with solvents (polar and not polar) and without solvent for the molecule and its ions (charge states -2, -1, +1, +2). In her we observe that the results obtained with PM3 and Spartan without solvent coincide practically for the neuter states and charge.
0
**Table 1**. Results of PM3 ( and Spartan for the PTCDA and NTCDA, in their neutral and charge states. We show the heats of formation (HF).

|   |   | Mopac | Spartan | | |   |   | Mopac | Spartan | | |
|---|---|---|---|---|---|---|---|---|---|---|---|
|   | HF Kcal/mol |   | No Solv. | No polar | water |   |   |   | No Solv. | No polar | water |
| P | Charge= -2 | -234.92 | -234.922 | -246.505 | -237.421 | N | -243.938 | -243.940 | -247.509 | -243.940 |
| T | Charge= -1 | -225.075 | -230.093 |   |   | T | -259.422 | -235.603 |   |   |
| C | Charge= 0 | -147.661 | -147.662 | -163.827 | -158.141 | C | -185.823 | -185.814 | -194.675 | -185.814 |
| D | Charge= +1 | 58.190 | 48.566 |   |   | D | 47.995 | 51.856 |   |   |
| A | Charge= +2 | 339.003 | 339.001 | 317.270 | 317.101 | A | 380.711 | 380.069 |   | 363.098 |

Analyzing the heat of formation of the structure first in the two methodologies is seen that the PTCDA and NTCDA are stable structures. It is particularly, doubly ionized in their negative and unstable in their positive charge state. It is also observed that the values of these heats of formation in the case of the molecules in neutral and ionic states (-2, +2) they spread to be smaller when are in presence of a solvent it is or not polar. There is an energetic gain for the molecular forms with the capture of one electron; the capture of a pair of electrons is energetically favorable to PTCDA. All results for the unpaired-electron systems because an approximation the stability of the anions is at worst underestimated.

The energetic cost to remove one electron is relatively high (~200 kcal/mol), and the cost of removing a pair of electrons is definitively prohibitive (~480 kcal/mol). In other words, this suggests that the PTCDA and NTCDA are strong electron acceptors.

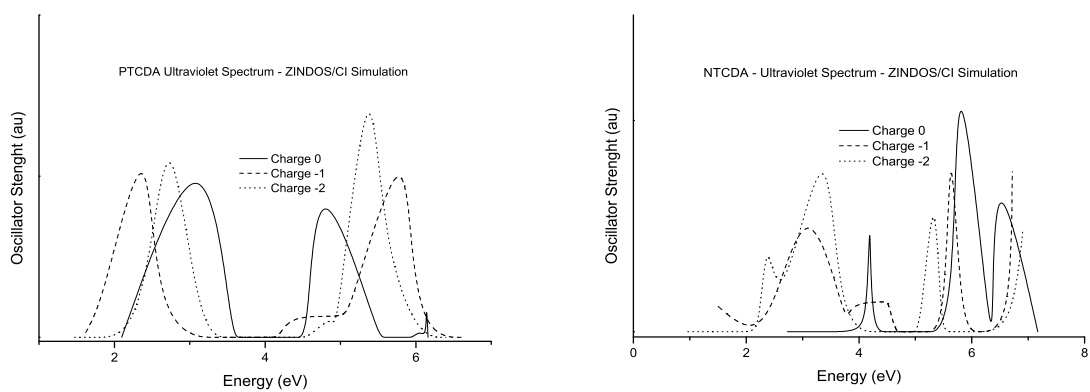

**Figure 2** Simulated optical absorption spectra of the PTCDA, NTCDA and their negative ions (-1, -2). These curves were obtained from the optical transition spectrum enveloped by normalized Gaussian functions weighted by the calculated oscillatory strengths.

It is of considering that most of the p-conjugated structures (plane) they are good donors of electrons (or, in semiconductors language, they are conductive for holes [5]), our results suggest a strong acceptor activity. This is, the negative PTCDA could present the appearance of a stable radical in the sense that the state is localized and the spin is one-half (S=1/2). In the figure 2, we show the simulated optical absorption spectra for the PTCDA, NTCDA and their negative ions (-1 and -2). As we can see from these figures in the neutral state the optic absorption it begins around the ~2.09 eV and 2.72 eV to Ptcda and Ntcda respectively.

These results are in good agreement with the experimental results that indicate that its gap this around the ~2.2 eV. But in the case of the ions the molecule begins to absorb around the 1.6/1.47 eV and 1.5/0.96 eV in their charge states, these results would indicate that the structures deposited experimentally can have a mixture of the neutral and ionized structures. The consideration that there is an energy gain under the electronic capture it reinforces the fact that the characterized structures

| System | Charge | First transitionn | Strongest Transition |
|---|---|---|---|
| PTCDA | -2 | 0.917 H→L+1> -0.254 H→L+3> | -0.960 H→L> 0.117 H→L+3> |
| | -1 | -0.928 H-1→H> 0.127 H-2→L> | -0.928 H-1→H> 0.127 H-2→L> |
| | 0 | -0.931 H→L> 0.169 H-2→L> | -0.931 H→L> 0.169 H-2→L> |
| NTCDA | -2 | 0983 H→L> -0.145 H-1→L+2> | 0987 H→L+1> 0.077 H-1→L+4> |
| | -1 | -0.923 H-1→L+1> 0.211 H-5→L> | 1.038 H-1→L+2> -0.347 H-1→L+3> |
| | 0 | 0.778 H→L+2> -0.526 H-1→L> | 0.625 H-3→L> -0.471 H-2→L+3> |

**Table 2.** Most relevant CI expansion coefficients for fully optimized from PTCDA and NTCDA in their Neutral and Charge states. Contributions are labeled according to convention H=HOMO, L=LUMO, L+1 = next in energy to LUMO and so.

The table 2 is showing the summary of the main CI contributions to the absorption threshold (first optically active electronic transition) and the highest peak (strongest active electronic transition) from ZINDO calculations for the PTCDA and their negative ions. As we can see from the table, the transitions are dominated by few configurations involving frontier orbitals, and as mentioned before, these are π-type orbitals. Our results show that the PTCDA and NTCDA are a plane structures, what contributes in the delocalization of the π electrons and consequently in the decrease of the optic gap of energy, which still diminishes but when the structures are deposited in a parallel way an on another. The consideration that there is an energy gain under the electronic capture it reinforces the fact that the characterized structures are a mixture of neutral and ionized structures. The results obtained for the PTCDA and NTCDA (neutral and ions) they are in good agreement with the experimental data and they show their behavior like semiconductor.

**Acknowledgement**
This work has been carried out partly under the support of the VRI (vicerrectoría de investigaciones) of the Universidad del Cauca inside the project and the FACNED by economic support to conference assist.